%% file: sample-sigconf.tex
\documentclass[sigconf]{acmart}
\usepackage{multirow}
\newtheorem{problem}{Problem}
\newtheorem{example}{Example}

\AtBeginDocument{%
  \providecommand\BibTeX{{%
    \normalfont B\kern-0.5em{\scshape i\kern-0.25em b}\kern-0.8em\TeX}}}

\copyrightyear{2021}
\acmYear{2021}
\setcopyright{acmcopyright}\acmConference[WSDM '21]{Proceedings of the
Fourteenth ACM International Conference on Web Search and Data Mining}{March
8--12, 2021}{Virtual Event, Israel}
\acmBooktitle{Proceedings of the Fourteenth ACM International Conference on Web
Search and Data Mining (WSDM '21), March 8--12, 2021, Virtual Event, Israel}
\acmPrice{15.00}
\acmDOI{10.1145/3437963.3441762}
\acmISBN{978-1-4503-8297-7/21/03}

\begin{document}
\sloppy

\title{Temporal Meta-path Guided Explainable Recommendation}
\author{Hongxu Chen}
\affiliation{%
  \institution{University of Technology Sydney}
}
\email{Hongxu.Chen@uts.edu.au}

\author{Yicong Li}
\authornote{Having equal contribution with the first author}
\affiliation{%
  \institution{University of Technology Sydney}
}
\email{Yicong.Li@student.uts.edu.au}

\author{Xiangguo Sun}
\affiliation{%
  \institution{Southeast University}
}
\email{sunxiangguo@seu.edu.cn}

\author{Guandong Xu}
\authornote{Corresponding author}
\affiliation{%
  \institution{University of Technology Sydney}
}
\email{Guandong.Xu@uts.edu.au}

\author{Hongzhi Yin}
\authornotemark[1]
\affiliation{%
  \institution{The University of Queensland}
}
\email{h.yin1@uq.edu.au}

\begin{abstract}
  Recent advances in path-based explainable recommendation systems have attracted increasing attention thanks to the rich information provided by knowledge graphs. Most existing explainable recommendation only utilizes static knowledge graph and ignores the dynamic user-item evolutions, leading to less convincing and inaccurate explanations. Although there are some works that realize that modelling user's temporal sequential behaviour could boost the performance and explainability of the recommender systems, most of them either only focus on modelling user's sequential interactions within a path or independently and separately of the recommendation mechanism. In this paper, we propose a novel \textbf{\underline{T}}emporal \textbf{\underline{M}}eta-path Guided \textbf{\underline{E}}xplainable \textbf{\underline{R}}ecommendation (\textbf{TMER}), which utilizes well-designed item-item path modelling between consecutive items with attention mechanisms to sequentially model dynamic user-item evolutions on dynamic knowledge graph for explainable recommendations. Compared with existing works that use heavy recurrent neural networks to model temporal information, we propose simple but effective neural networks to capture users' historical item features and path-based context to characterise next purchased item. Extensive evaluations of TMER on three real-world benchmark datasets show state-of-the-art performance compared against recent strong baselines.\footnote{https://github.com/Abigale001/TMER}
  
\end{abstract}



\maketitle
\pagestyle{plain} 
\input{intro}
\input{problem-def}
\input{model}
\input{experiment}

\input{related-work}

\section{Conclusion}\label{sec:con}
We propose TMER, which explicitly models dynamic user-item interactions over time with path-based knowledge-aware explainable capabilities. We explore item-item paths between consecutive items with attention mechanisms for users' sequential behaviour modelling. Future works may include the following directions: 1) generate human-readable explanations for recommendation with NLP techniques, and 2) explore causality learning \cite{xu2020causality} to discover more appealing paths for explainablity.


\begin{acks}
\vspace{-0.5em}
This work is supported by the Australian Research Council (ARC) under Grant No. DP190101985, DP170103954, DP200101374 and LP170100891.
\end{acks}

\bibliographystyle{ACM-Reference-Format}
\bibliography{ref.bib}

\end{document}

%% file: intro.tex
\section{Introduction}
Reasoning with paths over user-item associated Knowledge Graphs (KGs) has been becoming a popular means for explainable recommendations \cite{xian2019reinforcement, wang2019explainable, zhao2020leveraging}. The path-based recommendation systems have successfully achieved promising recommendation performance, as well as appealing explanations via searching the connectivity information between users and items across KGs. One category of existing works on path-based explainable recommendations seeks auxiliary meta-paths (pre-defined higher-order relational compositions between various types of entities in KGs) as similarity measures and evidence for possible explanations between users and items. 

\begin{figure}[ht]
	\vspace{-0.35cm}
	\includegraphics[scale=0.274]{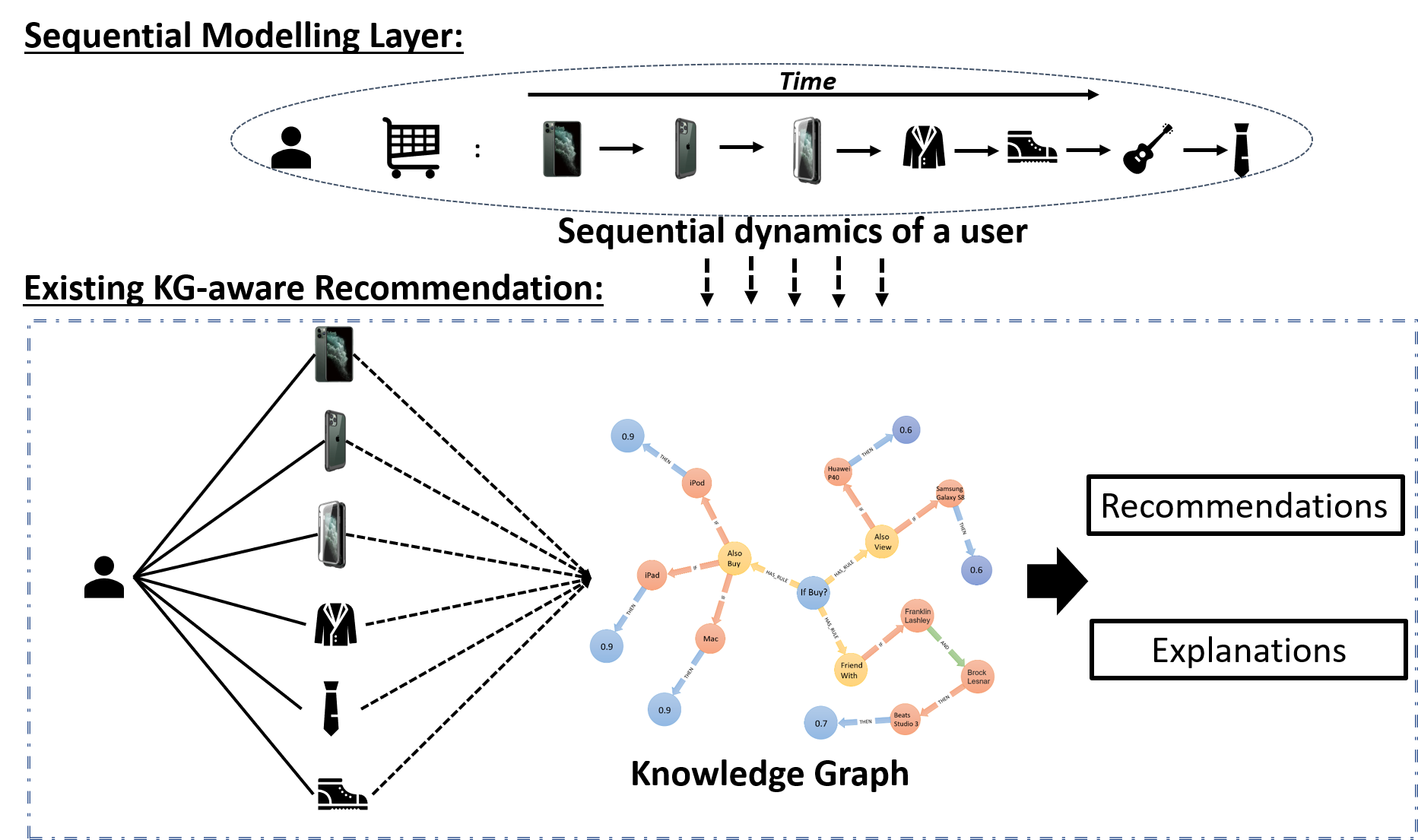}
	\caption{The concept of TEMR. TEMR contributes a sequential dynamic modelling layer on top of existing knowledge-aware explainable recommendation (the dashed box on the bottom).}
	\vspace{-0.5cm}
\end{figure}

However, most of existing path-based methods for explainable recommendation simply treat the underlying KGs as static graphs, ignoring the dynamic and evolving nature of user-item interactions in real-world recommendation scenarios. The dynamics and evolutions of users' interactions with items play a pivotal role in both recommendation precision and explanations for a users' real intent. Take the scenario in Figure 1 as an example, a user purchased a phone case and a phone film after a recent buy of a new phone. If we ignore the sequential information between each purchase and treat the whole information as a static graph, the system probably gives an explanation of buying the phone case is because a similar customer also bought this phone case by exploring co-purchasing relationships. Whilst the explanation, in this case, might be valid and the underlying reasoning mechanism (collaborative filtering signal) can be used for recommendation, it is still sub-optimal as a more appealing motivation of buying a phone case is due to the recent purchase of a new phone. For this reason, in this example, it is important for a system to be capable of considering temporal and sequential user-item interactions and disentangling the importance of various reasons when generating possible explanations. 

Although some prior works have considered some extent of the sequential information for knowledge-aware path-based explainable recommendation problem, they still fail to explicitly model the dynamics of users' activities. To allow effective reasoning on paths to infer the underlying rationale of a user-item interaction, the method proposed in \cite{wang2019explainable} takes the advantages of path connectivity and leverages the sequential dependencies of entities and sophisticated relations of a path connecting a user-item pair. Nevertheless, the methods only consider the sequential information within specific paths and did not consider the importance of the user' historical sequences on reflecting the user's dynamic interactions with items. To improve, the KARN model \cite{zhu2020knowledge} fuses the user's clicked history sequence and path connectivity between users and items in KGs for recommendation. However, the method models the user's sequential behaviours and user-item interactions separately and in a coarse-grained manner (treating a user's click history as a whole), which may restrict the expressiveness of users' temporal dynamics on recommendation explainability. 

In light of this, in this paper, we challenge the problem of exploring users' temporal sequential dynamics in the context of path-based knowledge-aware recommendation. Different from existing works that either only consider the sequential information within a path or treat the user's sequential interactions as a whole and separately, we aim to 1) explicitly model and integrate the dynamic user-item interactions over time into the path-based knowledge-aware recommendation, and 2) leverage the captured dynamics of user-item interactions to improve the performance and explainability of the recommendation .

It is worth noting that modelling users' temporal sequential behaviour with path-based knowledge-aware explainable recommendation is a non-trivial and challenging task. First, path-based knowledge-aware recommender systems are built upon the well-constructed KGs, and its expected accuracy and explainability are highly related to the underlying KGs and distilled paths. If temporal information between users and items is considered, the original underlying static KGs will be cast into multiple snapshots w.r.t. the timestamps. Comparing with the static KGs that consist of all users' full timelines, each snapshot at a certain timestamp only has partial observations of the global knowledge, which will result in inferior recommendation performance. To deal with the issue, we devise a general framework for temporal knowledge aware path-based explainable recommender systems, namely \textbf{Temporal Meta-path Guided Explainable Recommendation (TMER)}. TMER provides a solution that can explicitly depict users' sequential behaviour while being able to be aware of global knowledge of the entire underlying KGs. Specifically, to model the temporal dynamics between users and items, TMER naturally models the task as a sequential recommendation problem and takes as input users and their corresponding sequential purchase history. To learn the sequential dependencies between consecutive items purchased by a user, TMER novelly explores item-item paths between consecutive items and embed the paths as context with elaborated designed attention mechanisms to model the dynamics between user and items. Compared to existing path-based explainable recommendation systems that only consider user-item paths as evidence and support for the recommendation decision, TMER contributes another creamy layer on the top of existing works, which makes use of the powerful expressiveness of temporal information between users and items. 

In addition, when taking paths into consideration for the above purpose, another challenge lies in the existence of a large number of possible paths. It is time-consuming for a model to select several paths that are meaningful, expressive, and have positive impacts on both recommendation performance and explainability. To this end, inspired by prior work \cite{hu2018leveraging}, we leverage the concept of meta-path \cite{sun2011pathsim} and explore diverse meta-path schemas to characterize the context of dynamic interactions between users and items. With resort to the powerful transformer model for sequential modelling, we also elaborate item-item path attention and user-item path attention units to learn combinational features of multiple paths to further characterize users' temporal purchasing motivations and their general shopping tastes, respectively. The rationale for developing such path-based attention units is that a user's motivation towards buying a certain product is complex and consists of multiple factors. For example, when buying a new phone, a customer may consider several factors including a phone's intrinsic features such as functionality, display, camera, etc., as well as other external factors such as the choices of their close friends, and their previous purchase of certain related electronics using the same operating system. With help of the designed path-based attention units, the proposed TMER framework is able to learn different weights for various possible paths which will be then used as explanations for recommendations (we show the effectiveness of using the proposed path-based attention units for the explainable recommendation in the experiments by running case studies). 

Another shining point of our work is that the introduced item-item path modelling also serves as the core of the simple yet effective sequential modelling architecture of TMER. The combinational meta-path enriched features learned by the item-item path attention units are aggregated with features of previous purchased item will serve as the prediction signals for the next item. The intuition behind this framework is that the item-item paths connect two consecutive items purchased by a user, and represent various reasons and factors that may lead to the next purchased item, which intrinsically provides strong sequential dependencies between items. Compared to existing works on the sequential recommendations that rely on Recurrent Neural Nets such as (RNNs \cite{liu2016context}, GRUs \cite{hidasi2018recurrent}, LSTM \cite{wang2019explainable, zhu2020knowledge}), the proposed methods in this paper is light, simple but effective. 

The main contributions of this paper are summarized as follows:
\begin{itemize}
	\item We point out that explicitly modelling dynamic user-item interactions over time can significantly benefit the recommendation performance and explainability.
	
	\item We propose \textbf{T}emporal \textbf{M}eta-path Guided \textbf{E}xplainable \textbf{R}ecommendation (\textbf{TMER}), which considers users' dynamic behaviours on top of the global knowledge graph for sequential-aware recommendation and explores both user-item and item-item meta-path paths with well-designed attention mechanisms for explainable recommendation.
	
	\item Extensive evaluations on three real-world datasets have been conducted, and the experimental results showcase the superiority, effectiveness and temporal explainability of the proposed TMER model. 
	
\end{itemize}

%% file: problem-def.tex
\section{Preliminaries}


%
We first give some essential definitions and define the problem. 
\label{sec_Problem_Definition}
\begin{definition}
	\textbf{Information Networks.} An information network is a simple graph $G=(\mathcal{V},E)$. Each edge $e\in E $ represents a particular relation $r \in R$ of two entities$(v_1,v_2)$ linked by edge $e$. Each entity $v\in \mathcal{V}$ belongs to a particular type $T$. $(v_1,r,v_2)$ is a triplet means head entity $v_1$ to tail entity $v_2$ is with relation $r$. In general, relation $r$ and its reverse ${r}^{-1}$ is not the same unless the relation is symmetrical.
\end{definition}
\begin{definition}
	\label{pro_def_2.2}
	\textbf{Heterogeneous Information Network.} A heterogeneous information network (HIN) is a special type of information network. In heterogeneous information network $H$, edges $E$ and entities $\mathcal{V}$ are in different types, that is, $|R|>1, |T|>1$.
\end{definition}

\begin{figure}[ht]
	\centering
	\includegraphics[width=0.4\textwidth]{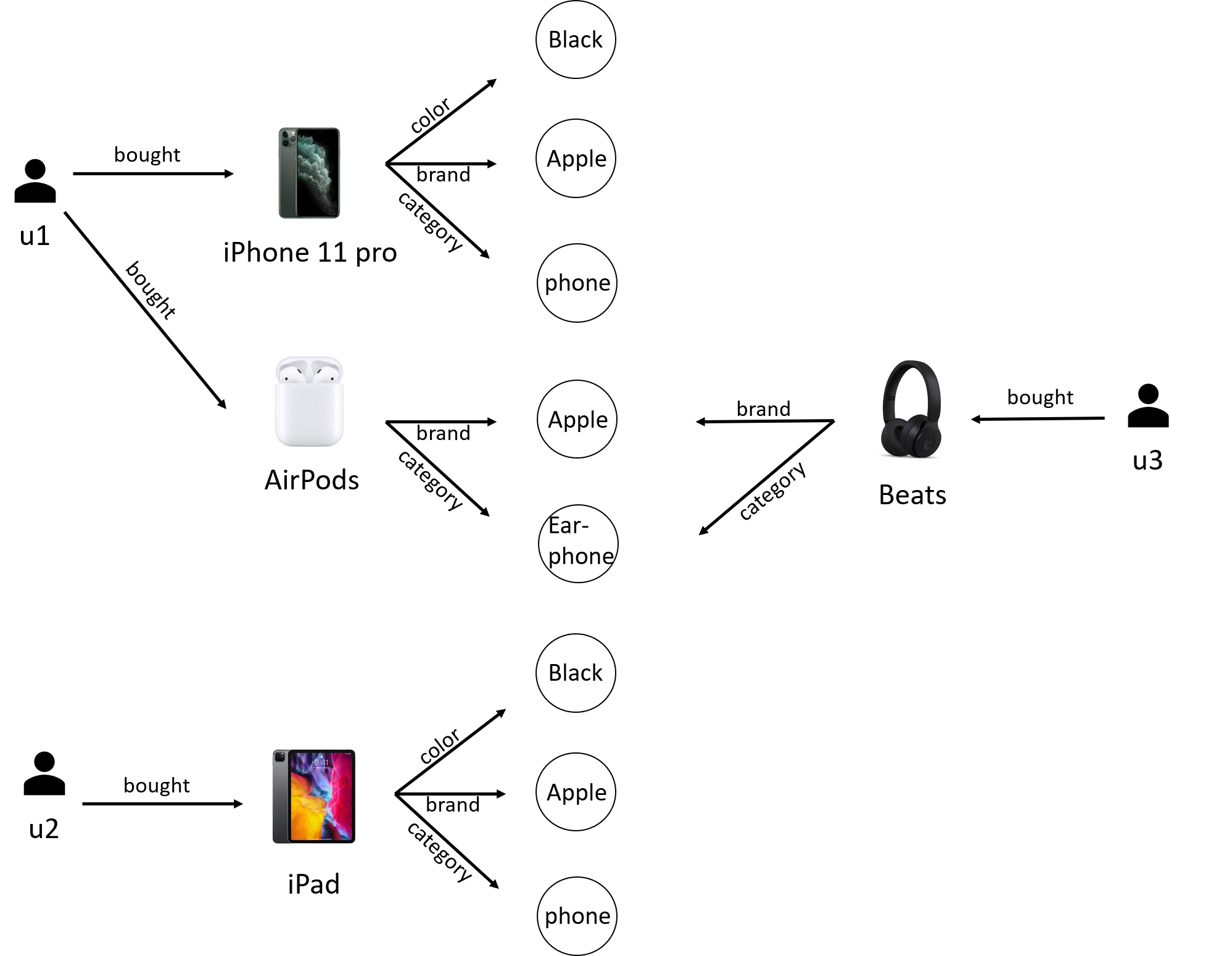}
	\vspace{-1em}
	\caption{An example of a heterogeneous information network: product recommendation network} \label{example_product}
	\vspace{-2em}
\end{figure}

\begin{example}
	As shown in Figure \ref{example_product}, A product recommendation network $G_P$ is a HIN, in which entities $\mathcal{V}$ could be many types, including items, users, categories, colors, brands, etc. $E$ could represent different relations. Edges between users and items denote the {\itshape buy} relation, edges between items and brands denote the {\itshape is brand of} relation, etc.
\end{example}

\begin{definition}
	\textbf{Meta-path}. A meta-path \cite{sun2011pathsim} $P$ in a network from entity $v_0$ to $v_k$, is denoted as $v_0\stackrel{r_0}{\longrightarrow}v_1\stackrel{r_1}{\longrightarrow}v_2\stackrel{r_2}{\longrightarrow}...\stackrel{r_{k-1}}{\longrightarrow}v_k$, where composite relation from $v_0$ to $v_k$ is $r=r_0 \circ r_1\circ r_2\circ ...\circ r_{k-1}$. $\circ$ represents the composition operator on relations.
\end{definition}
\begin{example}
	In the product recommendation network $G_P$ in figure \ref{example_product}, there are many meta-paths; one of the meta-paths is $UIBI$, which means $user \ {\rightarrow} \ item \ {\rightarrow} \ brand \ {\rightarrow} \ item$. In our paper, the chosen meta-paths will be discussed in the later sections.
\end{example}

\begin{problem}\textbf{Sequential knowledge-aware explainable recommendation.} For a user $u_i \in U$, given the item set $I$, the user transaction sequence $\pi_u$ and the item associated information network $G$, the target of knowledge-aware explainable recommendation is to predict top $k$ items that $u_i$ will interact with, as well as the possible reasonings of recommended items. 
\end{problem}

%% file: model.tex
\section{Model}
\begin{figure*}[h]
	\centering
	\includegraphics[width=0.8\textwidth]{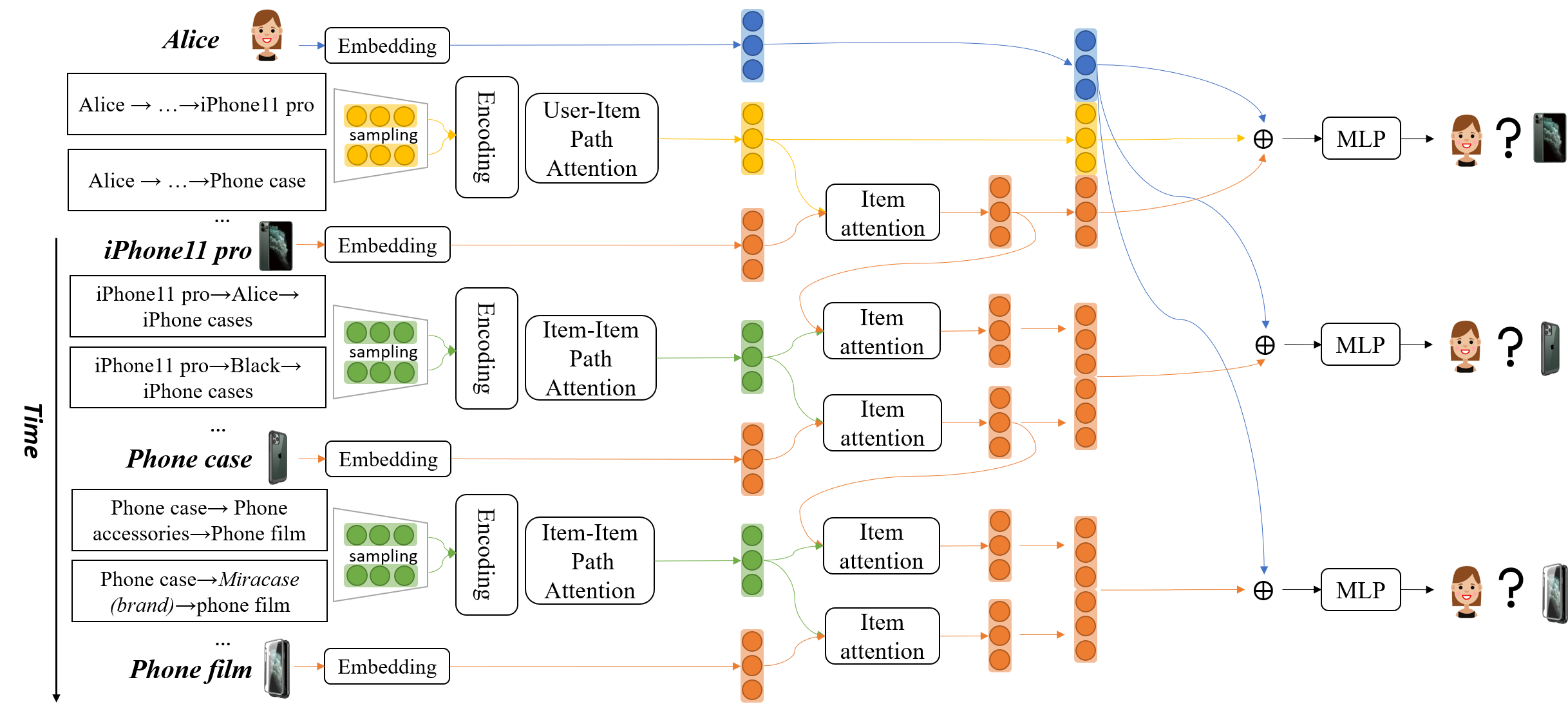}
	\vspace{-1em}
	\caption{The architecure of Temporal Meta-path Guided Explainable Recommendation. Here shows an example: User \textit{Alice} bought \textit{iPhone 11 pro}, \textit{Phone case} and \textit{Phone film} in a sequential order and the training process for \textit{Alice}.} \label{architecture}
	\vspace{-1em}
\end{figure*}
In this section, we introduce the proposed model Temporal Meta-path Guided Explainable Recommendation. In the remaining of the paper, we use the notation summarized in Table \ref{table_notations} to refer to the variables and parameters used throughout the paper.

\subsection{Overview of TMER Architecture}
The overall architecture of the proposed TMER model is shown in Figure \ref{architecture}. It mainly consists of four components. First, to initialize users and items, we use DeepWalk \cite{perozzi2014deepwalk} to embed user and item entities. Secondly, instead of simply utilizing meta-paths, we mine and extract both meaningful sequential (temporal) and non-sequential meta-paths instances to improve the recommendation performance and personalization. In this step, we obtain item-item instance paths between consecutive items. After embedding instances and user/item vectors with attention mechanism, we consider the weights of instances as the weights of reasoning paths for the specific user. We utilize different weights of reasoning paths to update item embeddings making items contain reasoning information. This step also models the users' sequential purchased information, feeding the previous item's feature to the next one. Finally, we feed item embeddings, user embeddings, and instances to the recommendation network to do recommendation. The specific steps are elaborated in the following subsections.

\begin{table}[t]
	\begin{tabular}{c|c}
		\hline
		\textbf{Symbol}  & \textbf{Description}                          \\ \hline
		$G$              & {Information network}    \\\hline
		$H$              & Heterogeneous information network (HIN)       \\\hline
		$e\in E$         & an edge                  \\\hline
		$v\in \mathcal{V}$         & an entity or node            \\\hline
		$R$            	 & Type of edges (relations) \\\hline
		$T$            	 & Type of nodes (entities)   \\\hline
		$p\in P$         & a meta-path in a HIN                              \\\hline
		$u\in U$         & User $u$                     \\\hline
		$i\in I$         & Item $i$                    \\\hline
		$W$              & Weight matrix                                 \\\hline
		$\textbf{b}$              & Bias vector                                   \\\hline
		$\phi \in \Phi$  & path instance in a HIN                               \\\hline
		$h$              & Embedding                                     \\\hline
		$\alpha$         & Attention mechanism parameter                 \\ \hline
		$\hat r$              & User-item rating                             \\ \hline
	\end{tabular}
	\caption{\label{table_notations}Description of major notations used in this paper}
	\vspace{-3em}
\end{table}

\subsection{Initialize User and Item Representations}
\label{subsec3.2}
We firstly learn latent representations for involved users and items by treating random truncated random walks in a user-item bipartite network as an equivalent of sentences in DeepWalk \cite{perozzi2014deepwalk}, which optimizes the co-occurrence probability among the entities in a walk by using skipgram based on word2vec \cite{mikolov2013efficient}. Other recent advanced graph-based embedding initialization methods can be also applied, like PME \cite{chen2018pme}, GraphSage \cite{graphsage}, GAT \cite{velivckovic2017graph}, and so on. These recent works usually can even outperform DeepWalk. However, through extensive comparison with these methods, DeepWalk is the best choice, the possible reason behind this is because DeepWalk pays more attention to the embedding of nodes in a path, while GCNs, such as GraphSage, GAT, etc., learn embedding of each entity with the aggregated feature information from its local neighborhoods. That is, these approaches focus on modelling local relationships. However, if only considering the recommendation task itself, although the local relationships are significant, the global higher-order relationships learned by walks on a graph also play a notable role. For example, the transitivity of co-purchasing relationships between friends and the substitute relationship of items can be discovered by modelling higher-order relations in a bipartite user-item graph. 

\vspace{-1em}
\subsection{Incorporating Meta-path based Context}
\label{subsec_3.3}
In recommendation tasks, external features related to users and items, such as product attributes \cite{chen2020try, xu2011semrec}, user social information \cite{chen2020social, yin2019social,gu2020enhancing}, and user demographic information are usually considered as additional auxiliary information to complement traditional recommendation methods. However, how to utilize the heterogenous additional information efficiently is an open problem. Some prior works \cite{jamali2010matrix, ma2008sorec, yuan2011factorization} attempted to consider social relations as the user-side information to boost the recommendation performance. 
To seek help from injecting more complex additional information, recent works \cite{yu2014personalized, han2018aspect} introduced meta-paths into recommendation methods to describe relational compositions between various types of entities in heterogeneous information networks. In \cite{yu2014personalized}, the authors proposed to diffuse user preferences along different meta-paths in information networks to generate latent features of users and items. Related work \cite{han2018aspect} firstly extracts different-aspect features with meta-paths from a HIN, and then fuse aspect-level latent factors to the recommendation systems. However, these methods largely rely on the latent factors obtained from constructed meta-path based similarity matrix, which are too general and only can reflect mutual interaction between different types of entities in a graph but cannot capture the specific information along particular path instances. Therefore, inspired by existing work \cite{hu2018leveraging}, we explore to improve both recommendation performance and explainability by modelling more specific meta-path instances. 


Different from existing works, we differentiate meta-path instances into two different categories based on the involved entities in a recommendation scenario (i.e., user-item and item-item meta-path instances). Through modelling these path instances, we learn a more detailed meta-path based context to further characterize the motivations, reasons as well as leading factors between each pair of user-item interaction. While previous works such as \cite{hu2018leveraging} mainly focuses on modelling meta-path instances between a user and an item, this paper highlights the item-item meta-path instances, which we think is beneficial in multiple aspects to sequential explainable recommender systems. Firstly, only considering user-item paths is restrictive for recommendation explainability as user-item paths only represent a user's general shopping interests. In comparison, item-item paths are more expressive and can reflect devise reasons by exploring higher-order relations among items, such as complementary products (e.g., phone and a phone case), substitutable items of known items (e.g., iPhone - phone film - Huawei Phone), co-purchased products with other people, etc. In addition, item-item paths sometimes also serve as sequential modelling signals that naturally capture the temporal dependencies between each consecutive item purchased by users, which will be of great impact for sequential explainable recommendations.

Despite the powerful expressiveness of meta-paths in exploring HIN based knowledge-aware recommendation, it is still challenging mainly because the number of meta-paths is too large to handle (i.e., the amount of edges is cubic to the number of entities). Taking the electric product recommendation situation for an example, for $IUIBI$ meta-path schema, if we fix the starting node - iPhone11 pro, there are many instances: $iPhone11 \ pro\rightarrow Alice\rightarrow iPad \rightarrow Apple\rightarrow AirPods$, $iPhone11 \ pro\rightarrow Amy\rightarrow Phone \ case \rightarrow Miracase \rightarrow Phone \ film$, and so on. Therefore, it is necessary to sample useful path instances while limiting the total amount to simplify later calculations. Intuitively, if the embedding of the next node and the current node is similar, the path to the next node is more likely to be connected. Therefore, according to this assumption, we use a method based on adjacent nodes' similarity to sample the next node along the path. Specifically, for each specified meta-path schema, we measure the priority degree by calculating the similarity between the current node and candidate out-going nodes. Such a priority score directly reflects the association degree between two nodes. Through this approach, we sample and get a few meta-path instances for each user-item pair and item-item pair to participate in recommendation module training. Taking meta-path $UIBI$ as an example, we firstly sample meta-path instances according to the similarity of $U$ and $I$, and $B$ and $I$ to get the start sub-paths and end sub-paths, getting top $k$ similar paths, and then, we calculate the similarity of $I$ and $B$ to get top $k$ middle paths to get all sampled paths. In what follows, we elaborate on how to embed the path-based context.

\vspace{-0.3em}
\subsection{Parameterizing Combinational Features of Meta-paths as Recommendation Context}
\label{subsec_seq_and_non_seq}
\subsubsection{\textbf{Learning combinational path-based features with Self-Attention}}
\label{subsubsec3.3.1}
After obtaining sampled user-item meta-path instances and item-item meta-path instances, we first regard paths as sentences, nodes as tokens in sentences, using Word2Vec \cite{mikolov2013efficient} method and $Mean(\cdot)$ operations to learn path embedding. Then, we employ multi-head self-attention units to learn the meta-path based context (the User-Item and Item-Item Path Attention modules shown in Figure 3). The rationale of deploying such self-attention units here is that because, after sampling, there are still multiple paths between each pair of item-item (or user-item) representing particular distinct reasons (reasoning paths); and we observe that the reasons for buying consecutive two items are not simply unique; rather, the reasons are more complex and likely a mixture of multiple different reasons. For example, the reasons for a customer to buy a phone case right after his/her previous purchase of a new phone are probably a mixture of 1) his/her friends who own a similar phone and bought this particular phone case, 2) the phone case is the most popular match for the purchased new phone, 3) the color of the phone case matches the customer's preference. The potential reasons can be more than the listed, and again they can be represented by using various meta-paths.

Based on this observation, self-attentive properties of the Transformer model \cite{vaswani2017attention}, we aim to learn the combinational features from multiple path instances to better characterize the complex reasons between each connected pair of entities in the KG. 
\begin{equation}
	\begin{split}
	&Attention (Q_{\phi},K_{\phi},V_{\phi})=Softmax(\frac{Q_{\phi}K_{\phi}^T}{\sqrt{d_k}})V_{\phi}, \\
	&MultiHead (Q_{\phi},K_{\phi},V_{\phi})=Concat (head_1,...,head_m)W^O , \\
	\end{split}
\end{equation}
where $ head_i $ is $ Attention (W_i^QQ_{\phi},W_i^KK_{\phi},W_i^VV_{\phi})x $. Query $Q$, key $K$ and value $V$ are self-attention variables associated with path $\phi$, and $W$ is the weight. $d_k$ is the dimensionality (here $d_k = 100$). $Concat(.)$ is the concatenation operation.

%


\subsubsection{\textbf{Modelling temporal dependencies with item-item meta-path instances}}
\label{sec:subsubsec3.4.2}
To learn the temporal dynamics of each user, the proposed TMER framework resorts to the above-mentioned item-item meta-path instances together with the well-designed architecture to capture the sequential dependencies between two consecutive items. Compared with most existing works on sequential recommendation \cite{wang2019explainable, zhu2020knowledge, xu2019recurrent} that utilize recurrent neural networks to encode the temporal effects between items in a user's interacted sequence, the proposed TMER bypasses the de-facto default deployment of RNNs or LSTMs that sometimes make the model even heavier. Specifically, the proposed framework novelly model the temporal dependencies between two items by capturing 1) the information passed from the previous item through an item-attention unit, 2) item-item connectivity through a specific sampled path instance. Notably, the information passed from the previous item is an attentive aggregation of previous item-item connectivity information. For example, in Figure 3, whether Alice will buy the phone film is modelled by considering 1) the information passed from the phone case (which includes the connectivity between iphone11 pro and the phone case), and 2) the paths between the phone case to the phone film. As a result, the long-range and short-term "memory" in a sequence can be captured, and the extent of the goodness of the long and short term memory can be influenced by the length of the modelled sequence in different scenarios with different datasets.

\subsubsection{\textbf{Updating item representations}}
After updating representations of user-item meta-path instances and item-item meta-path instances according to a multi-head self-attention mechanism, we employ a different kind of attention mechanism to update item representations. It is obvious that the current item is mostly related to the last one, which means it is better to add the last item's information to the current one to contain temporal information. Besides, the current item is also related to the instances that arrived at this item. Therefore, we perform a two-layer attention mechanism to update item representations. Mathematically,

\begin{equation}
\label{eq3.4.2.2}
\textbf{h}_i^{(1)}=ReLU(W_{i-1} \textbf{h}_{i-1} + W_{\phi _{i-1\rightarrow i}}^{(1)} \textbf{h}_{\phi _{i-1\rightarrow i}}+\textbf{b}_i^{(1)}) \odot \textbf{h}_{i-1}
\end{equation}
\begin{equation}
\label{eq3.4.2.3}
\textbf{h}_i^{(2)}=ReLU(W_i \textbf{h}_i+W_{\phi _{i-1\rightarrow i}}^{(2)} \textbf{h}_{\phi _{i-1\rightarrow i}}+\textbf{b}_i^{(2)}) \odot \textbf{h}_i
\end{equation}
where $\textbf{h}_i^{(1)}$ and $\textbf{h}_i^{(2)}$ mean the first and second layer output of the item attention module, respectively. $\textbf{h}_{i-1}$ is the last item's latent representation. $\phi _{i-1\rightarrow i}$ is the instance from the $(i-1)^{th}$ item to the $i^{th}$ item. $W$ and $\textbf{b}$ with different superscripts denote different variables' weights and bias. However, until now, here is a problem with the calculation of the first item, because there is no item before it. To solve this problem, we involve the user-item instance from user $u$ to the first item into the update of the first item, as Eq. \ref{eq3.4.2.4} shows. Actually, the instance from $u$ to the first item is really important in the recommendation, because it is the first item user has bought and most of the other bought items have a sequential relation with the first item to some extent. That is why we embed it into the first item. 
\begin{equation}
\label{eq3.4.2.4}
\textbf{h}_{i=1} = ReLU(W_{i} \textbf{h}_{i}+W_{\phi_{u\rightarrow i}} \textbf{h}_{\phi_{u\rightarrow i}}+\textbf{b}_{i}) \odot \textbf{h}_{i}
\end{equation}
where $\phi _{u\rightarrow i}$ represents the path from user $u$ to the first item.

\subsection{The Complete Recommendation Model} 
\label{subsec3.5}

Finally, we concatenate user, item and instances information (calculated in \ref{sec:subsubsec3.4.2}) to a vector according to Eq. \ref{eq3.5.1}, and get user-item prediction scores through Multilayer Perceptron (MLP) with explainability instances.

\begin{equation}
\label{eq3.5.1}
h_{u,i}=[\textbf{h}_u;\textbf{h}_i^{(1)};\textbf{h}_i^{(2)}]
\end{equation}
where $[;]$ means vector concatenation. Here $h_{u,i}$ denotes the explicit mutual vector of the user, item, and implicit mutual of user-item meta-path instances and item-item meta-path instances. For the first item, the concatenation operation is different because of the dimension problem, and therefore, for the first item related to each user, the vector fed in MLP involves user-item instance, mathematically,
\begin{equation}
h_{u,i=1} = [\textbf{h}_u;\textbf{h}_{\phi _{u\rightarrow 1}};\textbf{h}_{i=1}]
\end{equation}
After that, final user-item rating calculates as follows.
\begin{equation}
\label{eq3.5.2}
r_{u,i}=MLP(h_{u,i})
\end{equation}
where the MLP contains a two-hidden-layer neural network with ReLU as the activation function and sigmoid function as the output layer. According to \cite{hu2018leveraging, he2016deep}, neural network models can learn more abstractive features of data via using a small number of hidden units for higher layers, we employ a tower structure for the MLP component, halving the layer size for each successive higher layer. 

We use implicit feedback loss with negative sampling \cite{he2017neural, tang2015line} as the whole loss function:
\begin{equation}
\label{eq3.5.3}
loss_{u, i}=-E_{j \sim P_{n e g}}\left[\log \left(1-r_{u, j}\right)\right]
\end{equation}
where models the negative feedback drawn from the noise distribution $P_{n e g}$, which is a uniform distribution following \cite{hu2018leveraging}.

%% file: experiment.tex
\section{Experiment}
\subsection{Experiment Settings}
\subsubsection{Datasets}
We use three widely used datasets from the Amazon\footnote{http://jmcauley.ucsd.edu/data/amazon/} platform \cite{ni2019justifying}, including musical instruments, automotive, and toys and games dataset.  Each dataset includes brand and category as the metadata. More details are shown in Table 2.

We select the latest twelve items for each user and order these items by timestamps, and then we choose the first two items as bridge items, the next four items as training items and the rest as test items. We create the Heterogeneous Information Network using \textit{user}, \textit{item}, \textit{brand} and \textit{category} in datasets.At last, user-item meta-path instances and item-item meta-path instances are generated and sampled according to section \ref{subsec_seq_and_non_seq}.

\begin{table}[]\centering

	\caption{\label{Dataset Information}Dataset Information.}
	\vspace{-1em}
	\begin{tabular}{ccccccc}
		\hline
		\textbf{Datasets}& \textbf{User}& \textbf{Item} & \textbf{Brand} & \textbf{Category} \\ \hline
		Musical Instruments &1450&11457&1185&429\\ 
		Automotive &4600&36663&3790&1592\\ 
		Toys and Games&9300&58743&5404&820\\ \hline
	\end{tabular}
	\vspace{-1em}
\end{table}

\subsubsection{Evaluations}

We leverage Top $K$ Hit Ratio (HR@$K$) and Top $K$ Normalized Discounted Cumulative Gain (NDCG@$K$) as our metrics. For each positive test instance ($u, i^+$), we mix the ground truth item $i^{+}$ with $\mathcal{N}$ random negative items, then rank all these $\mathcal{N}+1$ items and compute the HR@$K$ and NDCG@$K$ scores. We set $\mathcal{N}=500$ and $K=1,5,10,20$ for evaluations.
For evaluation of the explainability of the recommendation, we use case studies to show the explanations in detail.

\subsubsection{Baselines}
We briefly introduce the baseline methods for comparison below.
\begin{itemize}
	\item \textbf{GRU4Rec}\cite{hidasi2015session, hidasi2018recurrent}: GRU4Rec is a session-based recommendation method using GRU. For each user, we treat the training items as a session. 

	\item \textbf{NARRE}\cite{chen2018neural}: NARRE utilizes neural attention mechanism to build an explainability recommendation system. 

	\item \textbf{MCRec}\cite{hu2018leveraging}: MCRec develops a deep neural network with the co-attention mechanism to learn rich meta-path based context information for recommendations. 


	\item \textbf{NFM}\cite{he2017neural2}: NFM effectively combines the linear Factorization Machines (FM) and nonlinear neural networks for prediction under sparse settings. 
		
	\item \textbf{FMG}\cite{zhao2017meta}: FMG uses FM and matrix factorization (MF) for recommendation and outperformed state-of-the-art FM and other HIN-based recommendation algorithms.
\end{itemize}
\vspace{-1em}
\subsection{Parameter Settings}
We choose the first second item of each user as the bridging item, the last 6 items of each user as the testing positive items, and the remaining 4 are training items. 
The hyperparameters are carefully tuned to achieve optimal results in all experiments. We implement GRU4Rec, NARRE and NFM based on their papers. The meta-paths and settings in MCRec are the same as the original paper. The meta-paths in FMG are \textit{User-Item (UI), User-Item-Brand-Item (UIBI)} and \textit{User-Item-Category-Item (UICI)}.
For our proposed TMER, the learning rate for Amazon Musical Instruments dataset is $5e-6$, for the Amazon Automotive dataset is $5e-5$ and for Amazon Toys and Games is $1e-4$. The parameters for other baselines are searched for their best results. We use \textit{UIBI, UICI, UIBICI} and \textit{UICIBI} as user-item meta-paths and \textit{ICIBI, IBICI, ICICI, IBIBI, IUIUI, ICIUI} and \textit{IBIUI} as item-item meta-paths. In these \textit{user/item/brand/category} related datasets, the selected meta paths are almost all the existing 4 to 6-hop meta paths.

\vspace{-1em}
\subsection{Effectiveness Analysis on Recommendation Results}

\begin{table*}[]
	\small
	\caption{\label{table:exp:baseline}Performance Comparison with Baselines.}
	\vspace{-1.5em}
	\resizebox{0.72\textwidth}{!}{
		\begin{tabular}{l|l|l|l|l|l|l|l}
			\multicolumn{1}{l|}{Datasets}             & Metrics & GRU4Rec & NARRE & MCRec & NFM    & FMG    & TMER   \\ \hline
			& HR@1    & 0.6493  & 0.6324 & 0.5922 &0.3929 & 0.4174 & \textbf{0.8467} \\
			& HR@5    & 0.6493  & 0.6408 & 0.6289 &0.4169 & 0.4328 & \textbf{0.9507} \\
			& HR@10   & 0.6502  & 0.6764 & 0.6522 &0.4631 & 0.6466 & \textbf{0.9739} \\
			& HR@20   & 0.7647  & 0.7988 & 0.6798 &0.5484 & 0.8799 & \textbf{0.9865}  \\ \cline{2-2}
			& NDCG@1  & 0.0490  & 0.1025 & 0.1584 &0.0549 & 0.1003 & \textbf{0.2165} \\
			& NDCG@5  & 0.0596  & 0.1224 & 0.1590 &0.0736 & 0.1221 & \textbf{0.2316} \\
			& NDCG@10 & 0.0632  & 0.1572 & 0.1590 &0.1013  & 0.1221 & \textbf{0.2432} \\
			\multirow{-8}{*}{Amazon Musical Instruments} & NDCG@20 & 0.0664  & 0.1603 &0.1592
			& 0.1432 & 0.1597 & \textbf{0.2614} \\ \hline
			& HR@1    & 0.6537  & 0.6633 & 0.6360 &0.3362 & 0.4032 & \textbf{0.9072}  \\
			& HR@5    & 0.6537  & 0.7063 & 0.6385 &0.4063 & 0.4645 & \textbf{0.9634} \\
			& HR@10   & 0.6576  & 0.7064 & 0.6412 &0.6024 & 0.8324 & \textbf{0.9697} \\
			& HR@20   & 0.8484  & 0.7963 & 0.6487 &0.6203 & 0.9596 & \textbf{0.9751} \\ \cline{2-2}
			& NDCG@1  & 0.0440  & 0.3343 & 0.1984 &0.1549 & 0.3242 & \textbf{0.4064} \\
			& NDCG@5  & 0.0642  & 0.4024 & 0.2080 &0.2742 & 0.3241 & \textbf{0.4691} \\
			& NDCG@10 & 0.0743  & 0.4536 & 0.2138 &0.3527 & 0.3360 & \textbf{0.4953} \\
			\multirow{-8}{*}{Amazon Automotive}        & NDCG@20 & 0.0773  & 0.4761 &0.2218& 0.3533 & 0.3897 & \textbf{0.5253} \\ \hline
			& HR@1    & 0.7776  & 0.8235 & 0.6580 &0.3563 & 0.4562 & \textbf{0.8339} \\
			& HR@5    & 0.7777  & 0.8644 & 0.6619 &0.4794 & 0.5536 & \textbf{0.9442} \\
			& HR@10   & 0.7792  & 0.8933 & 0.6668 &0.7335 & 0.8704 & \textbf{0.9580} \\
			& HR@20   & 0.8882  & 0.9025 & 0.6754 &0.7544 & 0.8978 & \textbf{0.9662} \\ \cline{2-2}
			& NDCG@1  & 0.1863  & 0.2065 & 0.1709 &0.0954 & 0.1124 & \textbf{0.2154} \\
			& NDCG@5  & 0.2046  & 0.2564 & 0.1800 &0.1543 & 0.1238 & \textbf{0.2901} \\
			& NDCG@10 & 0.2947  & 0.2933 & 0.1855 &0.2566 & 0.1462 & \textbf{0.3274} \\
			\multirow{-8}{*}{Amazon Toys and Games}               & NDCG@20 & 0.2947  & 0.3364 &0.1928& 0.2594 & 0.1514 & \textbf{0.3711}
	\end{tabular}}
	\vspace{-.5em}
\end{table*}

As shown in Table \ref{table:exp:baseline}, compared with these existing state-of-the-art recommendations, our proposed TMER achieves the best scores.
In terms of the hit ratio, on toys and games dataset, we improve at least 1\%, 8\%, 6\%, 6\% to the second when $K=\{1,5,10,20\}$, respectively. For NDCG, we outperform 0.9\%, 3\%, 6\%, 3\% to the second when $K=\{1,5,10,20\}$, respectively. The advantages also hold on other datasets. These results demonstrate that our framework can achieve obvious advantages through explicitly modelling each user's sequential purchased information with the temporal paths, while GRU4Rec \cite{hidasi2015session, hidasi2018recurrent} ignores the relation along paths and only utilizes RNN to model session-based recommendation. NARRE \cite{chen2018neural}, NFM \cite{he2017neural2} and FMG \cite{zhao2017meta} ignore the sequential information for each user. MCRec \cite{hu2018leveraging} ignores the item-item intrinsic relations and just utilizes user-item interactions to train the recommendation.
\begin{table}[]
	\vspace{-1em}
	\caption{\label{tab:exp:abl}Impact of User-Item and Item-Item Instances Self-Attention Mechanisms.}
	\vspace{-1em}
	\small
	\resizebox{0.45\textwidth}{!}{
		\begin{tabular}{l|l|c|c|c}
			{\color[HTML]{000000} Components}                                                                                 & {\color[HTML]{000000} Metrics} & {\color[HTML]{000000} TMER}   & {\color[HTML]{000000} TMER-RUI} & {\color[HTML]{000000} TMER-RII} \\ \hline
			{\color[HTML]{000000} }                                                                                           & {\color[HTML]{000000} HR@1}    & \textbf{0.8733} & {\color[HTML]{000000} 0.7844}                   & {\color[HTML]{000000} 0.7757}                   \\
			{\color[HTML]{000000} }                                                                                           & {\color[HTML]{000000} HR@5}    & \textbf{0.9541} & {\color[HTML]{000000} 0.8518}                   & {\color[HTML]{000000} 0.8374}                   \\
			{\color[HTML]{000000} }                                                                                           & {\color[HTML]{000000} HR@10}   & \textbf{0.9742} & {\color[HTML]{000000} 0.8606}                   & {\color[HTML]{000000} 0.8499}                   \\
			{\color[HTML]{000000} }                                                                                           & {\color[HTML]{000000} HR@20}   & \textbf{0.9871} & {\color[HTML]{000000} 0.8733}                   & {\color[HTML]{000000} 0.8641}                   \\ \cline{2-2}
			{\color[HTML]{000000} }                                                                                           & {\color[HTML]{000000} NDCG@1}  & \textbf{0.1688} & {\color[HTML]{000000} 0.0937}                   & {\color[HTML]{000000} 0.0947}                   \\
			{\color[HTML]{000000} }                                                                                           & {\color[HTML]{000000} NDCG@5}  & \textbf{0.2128} & {\color[HTML]{000000} 0.1593}                   & {\color[HTML]{000000} 0.1500}                     \\
			{\color[HTML]{000000} }                                                                                           & {\color[HTML]{000000} NDCG@10} & \textbf{0.2418} & {\color[HTML]{000000} 0.1938}                   & {\color[HTML]{000000} 0.1778}                   \\
			\multirow{-8}{*}{{\color[HTML]{000000} \begin{tabular}[c]{@{}l@{}}Amazon \\ Musical \\ Instruments\end{tabular}}} & {\color[HTML]{000000} NDCG@20} & \textbf{0.2817
			} & {\color[HTML]{000000} 0.2393}                   & {\color[HTML]{000000} 0.2147}                   \\ \hline
			{\color[HTML]{000000} }                                                                                           & {\color[HTML]{000000} HR@1}    & \textbf{0.9072} & {\color[HTML]{000000} 0.9060}                    & {\color[HTML]{000000} 0.8942}                   \\
			{\color[HTML]{000000} }                                                                                           & {\color[HTML]{000000} HR@5}    & \textbf{0.9634} & {\color[HTML]{000000} 0.9633}                   & {\color[HTML]{000000} 0.9421}                   \\
			{\color[HTML]{000000} }                                                                                           & {\color[HTML]{000000} HR@10}   & \textbf{0.9697} & {\color[HTML]{000000} 0.9693}                   & {\color[HTML]{000000} 0.9474}                   \\
			{\color[HTML]{000000} }                                                                                           & {\color[HTML]{000000} HR@20}   & \textbf{0.9751} & {\color[HTML]{000000} 0.9748}                   & {\color[HTML]{000000} 0.9531}                   \\ \cline{2-2}
			{\color[HTML]{000000} }                                                                                           & {\color[HTML]{000000} NDCG@1}  & \textbf{0.4064} & {\color[HTML]{000000} 0.3521}                   & {\color[HTML]{000000} 0.3694}                   \\
			{\color[HTML]{000000} }                                                                                           & {\color[HTML]{000000} NDCG@5}  & \textbf{0.4691} & {\color[HTML]{000000} 0.4456}                   & {\color[HTML]{000000} 0.4472}                   \\
			{\color[HTML]{000000} }                                                                                           & {\color[HTML]{000000} NDCG@10} & \textbf{0.4953} & {\color[HTML]{000000} 0.4794}                   & {\color[HTML]{000000} 0.4770}                    \\
			\multirow{-8}{*}{{\color[HTML]{000000} \begin{tabular}[c]{@{}l@{}}Amazon\\ Automotive\end{tabular}}}              & {\color[HTML]{000000} NDCG@20} & \textbf{0.5253} & {\color[HTML]{000000} 0.5155}                   & {\color[HTML]{000000} 0.5104}                  
	\end{tabular}}
	\vspace{-1em}
\end{table}

\vspace{-1em}
\subsection{Effectiveness Analysis on Attention}
To study the impact of user-item and item-item instances, we further compare our model with two variants, namely TMER-RUI, and TMER-RII. TMER-RUI means we do not consider user-item instance and remove the corresponding self-attention module. TMER-RII means we remove the self-attention module for item-item instances. We report the compared results in Table \ref{tab:exp:abl}, from which we can see that the user-item and item-item instances self-attention modules significantly boost the recommendation method, adaptively adjust the importances of different instance paths. The self-learned user-item and item-item instance paths contribute to improving the performance of recommendation finally. Especially, using the item-item meta-path instances self-attention module helps TMER improve 6.7\% and 12.3\% on Musical Instruments dataset on HR@$20$ and NDCG@$20$, respectively.

\begin{figure*}[htp]
	\centering
	\includegraphics[width=0.9\textwidth]{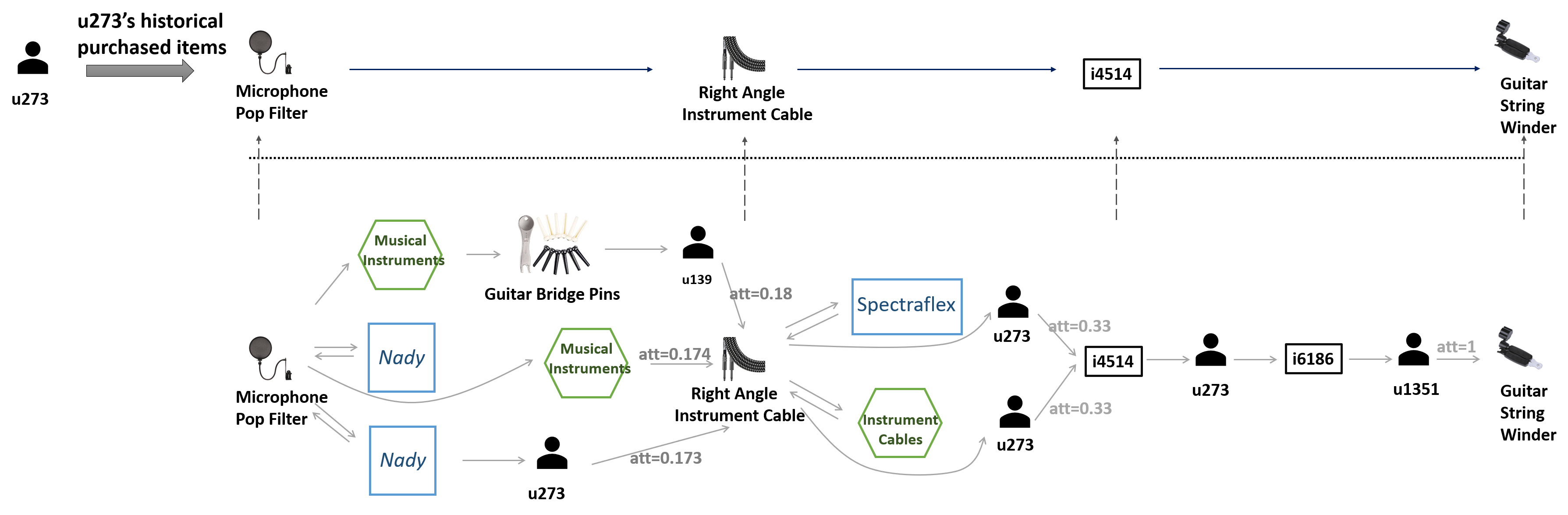}
	\vspace{-1em}
	\caption{It shows $u273$ and related historical purchased items in the upper part of the figure. In the lower part, it displays the reasoning item-item paths along this historical purchased items related to $u273$. Green blocks represent the category attributes. Blue blocks represent the brand attributes. Black blocks without physical pictures do not have meta information in the dataset.} \label{case_study}
	\vspace{-1em}
\end{figure*}

\vspace{-1em}
\subsection{Effectiveness Analysis on meta-paths}
\begin{figure}[htp]
	\centering
	\includegraphics[width=0.45\textwidth]{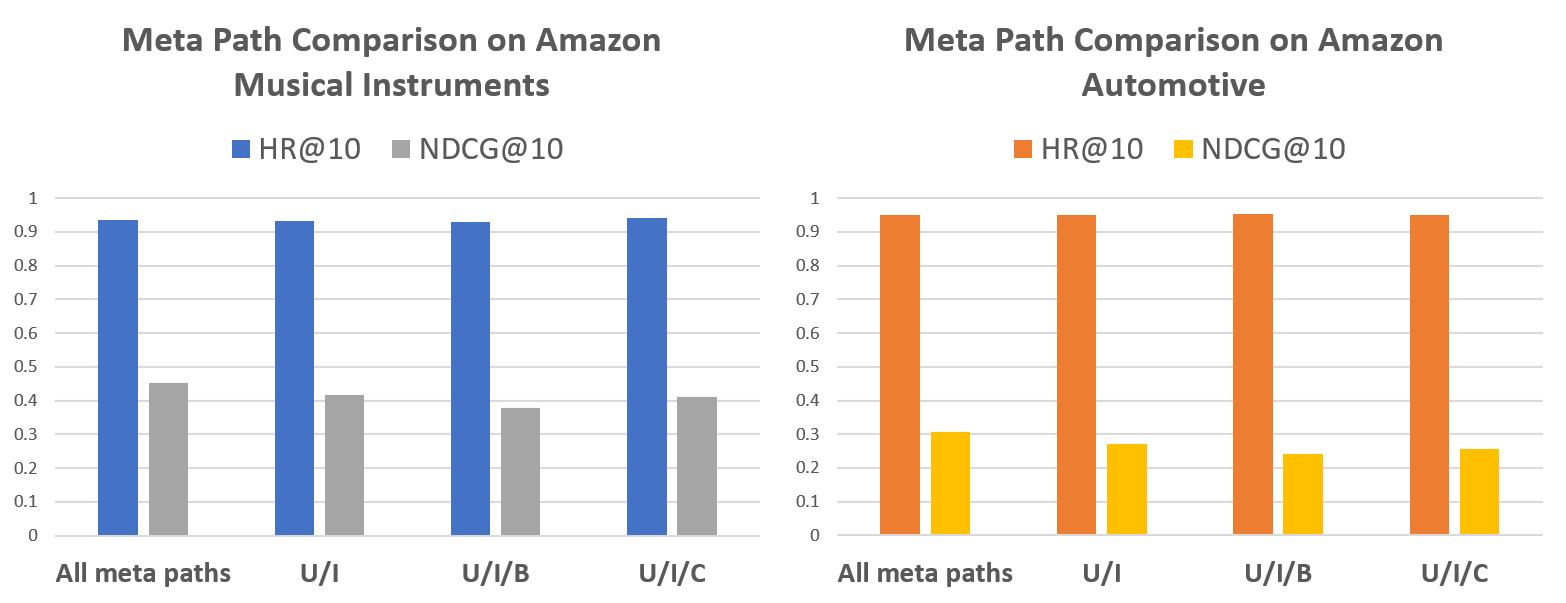}
	\vspace{-1em}
	\caption{Impacts of different meta-paths on TMER. The x-axis and y-axis represent different meta-path types and the value of evaluation metrics, respectively.} \label{fig:exp:meta_path_comparison}
\vspace{-.5cm}
\end{figure}

To analyze the contributions of different meta-path schemes, we randomly sample 100 negative items for each user-item positive pair and the evaluation metrics are HR@$10$ and NDCG@$10$. In Figure \ref{fig:exp:meta_path_comparison}, the four sets of bars denote results of all meta-paths, meta-paths only related to users and items, meta-paths only related to users, items and brands, meta-paths only related to the users, items and categories, respectively. We find the blue and orange bars (HR@$10$ on Musical Instruments and Automotive datasets) are almost the same. However, we can get the conclusion from the gray and yellow bars (NDCG@$10$ on Musical Instruments and Automotive datasets) that meta-path related to all attributes outperform the others. Moreover, from the trend of gray or orange bars, we can find that the meta-path related to \textit{category} improves more than \textit{brand}. 

\vspace{-1em}
\subsection{Case Study: generating explanation for recommendation}
One contribution of this paper is to give explanations on instances paths while recommending preferable items. This is because our method generates multiple item-item instance paths according to user purchase history, and then it utilizes attention mechanism to learn the weight of each item-item instance path and aggregate multiple item-item instance paths for each item-item pair. To demonstrate this, we show a random example drawn from TMER on Amazon Musical Instruments dataset.

We randomly select a user, whose id is $u273$, and track his item-item instance paths' parameters. In the training dataset, $u273$'s purchase history is $i2954$, $i2280$, $i4514$ and $i11158$. We shows several sampled item-item instance paths with high attention parameters in Figure \ref{case_study} and demonstrates our explanations.

\begin{itemize}
	\item According to Figure \ref{case_study}, there are three reasons for purchasing $i2280$. The most probable reason with the highest attention weight is that $u273$ bought the last musical instrument category $i2954$ and $i2237$ with the attention weight $0.18$. For the next item $i4514$, the reason for purchasing it is that the user $u273$ has bought $2280$ who has the same brand and category with item $i4514$. There is only one item-item instance path between some items because the item-brand and item-category data are sparse. Thus, our method can infer the reasons through the weights on different item-item paths.

	\item Besides, our model can capture sequential information according to user purchase history thanks to item-item paths. These item-item paths learn the reason path from the current item to the next item. In the whole model, these reason paths will feed to the item attention module. Therefore, our model can recommend with learned sequential information.
\end{itemize}

%% file: related-work.tex
\vspace{-1em}
\section{Related work}
\label{sec_related_work}
Early recommendation systems mostly rely on Collaborative Filtering (CF) \cite{schafer2007collaborative, sarwar2001item}, which are based on the idea that users with similar history will be more likely to purchase similar items. However, CF-based recommendations always suffer from sparsity issues and cold-start problems. Therefore, some works utilize side information, like user and item attributes \cite{gong2009employing}, item contents \cite{melville2002content, wang2018sequence} and other external information \cite{yin2017spatial} to solve these issues. Among them, knowledge graph-based methods \cite{wang2020reinforced, xian2019reinforcement, zhu2020knowledge, wang2019explainable} show great advantages to recommendation performance and explainability. 

Knowledge-aware recommendations can be roughly categorized into embedding-based and path-based approaches. Prior efforts on embedding-based methods \cite{zhang2016collaborative} always use embeddings of the knowledge graph to model the user-item interactions for recommendations. For example, exiting works \cite{huang2018improving} utilizes TranE \cite{bordes2013translating} to embed user-item interactions to integrate knowledge into the recommendation system. Similarly, \cite{grad2017graph} embeds users and items into the same embedding space for recommendation. The above approaches model the relations among users and items using knowledge graph embedding methods, which achieved great improvements in model expressiveness. However, these methods are sensitive to the quality of related knowledge graphs.

To this end, meta-paths \cite{sun2011pathsim}, and the connectivity among different types of nodes are introduced to recommendations, which have the ability to learn the explicit expression along each meta-path schema. In \cite{zhao2017meta}, it introduces Matrix Factorization (MF) and Factorization Machine (FM) to learn similarities generated by each meta-path for recommendation. \cite{han2018aspect} models rich objects and relations in knowledge graph and extracts different aspect-level similarity matrices thanks to meta-paths for the top-N recommendation. Although they achieved appealing performance, the limitation is still obvious that structured meta-path based similarity latent factors can only reflect mutual infectivity along meta-path schemas in a graph but cannot capture the specific information along particular path instances, which limits the explainability of recommendation.

More recently, injecting meta-paths as recommendation context (aggregation of meta-path instances) \cite{hu2018leveraging} was introduced for top-N recommendation. It provides fine-grained explanations based on specific instances. However, it ignores the important sequential dynamics of user-item interactions, limiting the performance of the recommendation performance and interpretability. To model sequentially, \cite{wang2019explainable} utilizes LSTM to model the sequential information, but it only sequentially models path between users and items and ignores the importance of the user's clicked history sequences, which are highly informative to infer user's preferences. To tackle the issue, \cite{zhu2020knowledge} attempts to model the sequences of user's behaviours and path connectivity between users and items for recommendation. Nevertheless, it only models user-item paths, which ignores the item-item intrinsic relation information and cannot learn complex semantic information between items. Based on the above research, we propose a meta-path based recommendation with differentiated user-item and sequential item-item instances to enhance the learning ability for explainability recommendation.